\documentclass[sigconf]{acmart}
\newif\ifreb
\rebfalse     

\usepackage{xcolor} 
\usepackage[normalem]{ulem} 

\newcommand{\rebadd}[1]{\ifreb\textcolor{red}{#1}\else#1\fi}
\newcommand{\rebdel}[1]{\ifreb\textcolor{red}{\sout{#1}}\else\relax\fi}
\newcommand{\rebrep}[2]{\rebdel{#1}\rebadd{#2}} 
\usepackage{threeparttable}
\usepackage{newunicodechar}
\usepackage{threeparttable} 
\usepackage{placeins}      
\usepackage{graphicx}  
\usepackage{booktabs}
\usepackage{array}
\usepackage{multirow}
\usepackage{tabularx}
\usepackage{enumitem}
\usepackage{caption}  
\usepackage{longtable}
\usepackage{tabularx}
\usepackage{listings}
\usepackage{xcolor}

\graphicspath{{figs/}{images/}} 
\newunicodechar{α}{\ensuremath{\alpha}}
\newunicodechar{β}{\ensuremath{\beta}}
\newunicodechar{γ}{\ensuremath{\gamma}}
\AtBeginDocument{%
  }
\usepackage{etoolbox}

\makeatletter
\newcommand{\bibitemcolor}[1]{%
  \ifreb
    \ifcsname bibitem@#1\endcsname
      \csname bibitem@#1\endcsname
    \else
      black%
    \fi
  \else
    black%
  \fi
}

\pretocmd{\@bibitem}{\color{\bibitemcolor{#1}}}{}{}
\pretocmd{\@lbibitem}{\color{\bibitemcolor{#2}}}{}{}
\makeatother

\newcommand{\highlightbib}[1]{%
  \expandafter\def\csname bibitem@#1\endcsname{red}%
}

\copyrightyear{2026}
\acmYear{2026}
\setcopyright{cc}
\setcctype{by}
\acmConference[IUI '26]{31st International Conference on Intelligent User Interfaces}{March 23--26, 2026}{Paphos, Cyprus}
\acmBooktitle{31st International Conference on Intelligent User Interfaces (IUI '26), March 23--26, 2026, Paphos, Cyprus}
\acmPrice{}
\acmDOI{10.1145/3742413.3789171}
\acmISBN{979-8-4007-2281-3/26/04}

\begin{document}

\title{SmartWalkCoach: An AI Companion for End-to-End Walking Guidance, Motivation, and Reflection}

\author{Xianzhe Zhang}
\email{XianzheZhang50@gmail.com}
\orcid{0009-0009-7616-6123}
\affiliation{%
  \institution{School of Advanced Technology, Xi'an Jiaotong-Liverpool University}
  \city{Suzhou}
  \country{China}
}

\author{Mingxuan Hu}
\email{Mingxuan.Hu22@student.xjtlu.edu.cn}
\orcid{0009-0004-0448-6367}
\affiliation{%
  \institution{School of Advanced Technology, Xi'an Jiaotong-Liverpool University}
  \city{Suzhou}
  \country{China}
}

\author{Bufan Xue}
\email{Bufan.Xue24@student.xjtlu.edu.cn}
\orcid{0009-0001-0852-4912}
\affiliation{%
  \institution{School of Advanced Technology, Xi'an Jiaotong-Liverpool University}
  \city{Suzhou}
  \country{China}
}

\author{Erick Purwanto}
\authornote{Corresponding author.}
\email{Erick.Purwanto@xjtlu.edu.cn}
\orcid{0000-0001-6497-6721}
\affiliation{%
  \institution{School of Advanced Technology, Xi'an Jiaotong-Liverpool University}
  \city{Suzhou}
  \country{China}
}

\author{Thomas Selig}
\authornotemark[1]
\email{Thomas.Selig@xjtlu.edu.cn}
\orcid{0000-0002-2736-4416}
\affiliation{%
  \institution{School of Advanced Technology, Xi'an Jiaotong-Liverpool University}
  \city{Suzhou}
  \country{China}
}

\author{Daniel Yonto}
\email{Daniel.Yonto@xjtlu.edu.cn}
\affiliation{%
  \institution{Urban Planning and Design, Xi'an Jiaotong-Liverpool University}
  \city{Suzhou}
  \country{China}
}

\renewcommand{\shortauthors}{Zhang et al.}

\begin{abstract}
  We present SmartWalkCoach, a mobile AI companion that supports the full walking journey: from pre-walk planning to in-walk guidance through to post-walk reflection. Addressing a gap between map navigation and motivational coaching, SmartWalkCoach orchestrates three lightweight agents: (1) \emph{GeographyAgent} for conversational route curation from nearby points of interest and user preferences while delegating pathfinding to map APIs; (2) \emph{AccompanyAgent} for context-aware, just-in-time prompts that blend informational cues with relational encouragement; and (3) \emph{SummaryAgent} for concise reflection and next-step planning. This end-to-end, tool-using design aims to lower cognitive load in planning and sustain engagement and motivation during walking through delivering dynamic, cadence-aware interventions. We conducted an in-the-wild, two-period AB/BA crossover study (N=12), where each participant completed two comparable walks with counterbalanced conditions: Information-only versus Information+Motivation. Linear mixed models show that adding motivational, companion-like dialogue significantly improved outcomes: participants reported higher positive feelings and better user experience, with no evidence of carryover. Thematic analysis surfaced two design imperatives for mobile companions: supportive, relational expression and context-aware timing (e.g., avoiding high-load moments, intervening at fatigue/milestones). Our contributions are: (i) an end-to-end, tool-using agent architecture for everyday walking that reduces cognitive load during planning and accompaniment; (ii) a controlled field evaluation linking context-aware motivation to affect and UX gains; and (iii) actionable design guidance on expression, timing, and frequency for mHealth companions. We outline limitations and paths toward multimodal, voice-first companions, with adaptive personalization mechanisms.
\end{abstract}




\maketitle

\section{Introduction}

\rebadd{Walking is one of the foundational elements of healthy living, with its powerful benefits to both physical and mental health being well established~\cite{WHO2021PhysicalActivity,Morris1997WalkingToHealth}.}
 As health awareness expands, numerous applications have been created to promote physical activity. Currently, much of the existing AI/LLM work on walking or exercise support emphasizes in-walk navigation and route direction tasks~\cite{Latif2024_3P_LLM, Meng2024_LLM_Astar, Li2025_GridRoute, Tariq2025_RobustLLM_DynamicWaypoints, Huang2024GeoAgent, Chen2024_CanLLMsPlanPaths}. In parallel, within the health/fitness domain, conversational agents and chatbots have been studied for their ability to promote physical activity, engagement, and behavior change~\cite{Wang2025ChatbotExerciseInterventions, Oh2021SystematicReviewChatbotsPhysicalActivity, Jorke2024_SupportingPhysicalActivityLLM}. However, several challenges remain inadequately addressed in supporting the complete walking journey.

Firstly, a noticeable gap exists between the two aforementioned strands: systems focusing on navigation often lack integrated motivational support, while conversational agents for stimulating physical activity may not be deeply coupled with real-time, context-aware geospatial guidance. Furthermore, many current walking app designs lack \emph{universality}. Instead, they target specific user groups, such as keen walkers engaging in goal-oriented activities like hiking or racing (see~\cite{Raghuveer2022PACE} for a typical example). This focus may limit their effectiveness for everyday walking environments and discourages their uptake by a more diverse range of users which includes occasional walkers.

Secondly, in the pre-walk planning phase, users often find it difficult to formulate routes that truly reflect their preferences and needs. There is an inherent information gap between what users intuitively desire and the options presented by map systems. Many users have only partial knowledge of nearby points of interest (POIs) and struggle to express vague but nuanced preferences, such as wanting “a quiet, shaded path with occasional cafés”, and then match those desires with the vast candidate pool. Traditional recommender systems, which rely heavily on rich historical interaction data, tend to suffer in scenarios involving new users or loosely specified preferences (i.e. cold‑start conditions). The work ZeroPOIRec is one example of using LLMs to provide zero-shot POI recommendations in such low-data contexts \cite{Kim2025ZeroPOIRec}. More recently, GA‑LLM (Geography-Aware LLM) demonstrates how language models augmented with spatial encoding and POI relationship reasoning can improve next-POI prediction performance under sparse data \cite{Liu2025GA-LLM}. Still, applying these advances specifically to walking route planning—where recommendations must jointly satisfy spatial constraints, personal preferences, and real‑world accessibility—remains a challenging problem.

Finally, another difficulty lies during the walk itself: walkers frequently experience motivational dips, which can lead to incomplete routes \cite{Andre2024BehavioralPerspectiveExerciseAdherence}. Current motivation mechanisms within apps are often restricted to passive navigation guidance or rudimentary gamification strategies, such as earning badges for distance covered~\cite{Shameli2017GamificationWalking}. While gamification elements can provide initial excitement, they often fail to sustain intrinsic motivation over the long term~\cite{Agarwal2021Gamification}. This approach overlooks the potential for context-aware, real-time companionship, which HCI and agent research suggests is crucial for establishing lasting, trusting relationships between users and computational systems. For instance, relational agents designed for long-term interaction have been shown to be more trusted, respected, and liked compared to task-only agents in month-long trials~\cite{Bickmore2005LongTermRelations}. Broader trust models in human-computer interaction emphasize the importance of predictability, consistency, and social cues in building trust over time~\cite{Gulati2024_TrustHCI, Daronnat2021_InferringTrust}. Studies in real-time collaboration contexts further indicate that agent behaviors directly influence users' reliance and confidence in the system~\cite{Daronnat2020HumanAgentTrust}. An ideal walking companion should therefore not only provide timely information but also deliver motivational support tailored to the user's current physiological and contextual state. This aligns with the principles of Just-In-Time Adaptive Interventions (JITAIs), which advocate for delivering the right type of support at the opportune moment to maximize behavioral impact~\cite{Hardeman2019Systematic, Klasnja2015MRT}.

We therefore ask the following research questions:
\begin{description}
\item[RQ1:] How can conversational interfaces for route planning reduce cognitive load and improve user satisfaction in mobile applications?
\item[RQ2:] In real-world walking, does an Informative + Motivational agent improve motivation, engagement, and satisfaction compared to an Informative-only agent?
\item[RQ3:] How to optimize the expression/frequency/timing of accompanying agents during walking?
\end{description}

To address these, we conceived and implemented \emph{SmartWalkCoach}, an Android-based program that realizes an end-to-end intelligent user interface for walking companionship empowered by AI.  SmartWalkCoach brings together three fundamental functional agents working together to oversee the whole walking experience. The \emph{GeographyAgent} is responsible for conversational route planning by employing geographical API tools alongside solicited user inclinations~\cite{Huang2024GeoAgent,Singh2024GeoLLMEngine,Kim2025ZeroPOIRec}. The \emph{AccompanyAgent} brings in-situ companionship with contextualized motivational encouragement and map information prompts~\cite{Sage2025Companionship,Luo2021PromotingPAAgents}. Lastly, the \emph{SummaryAgent} organizes post-walk reflection and consolidation and optional social media sharing~\cite{Sheeran2025WhenHow}.

We then conducted an in-the-wild, walk-along, experimental study to validate the effectiveness of SmartWalkCoach, with 12 users. Each user completed two walks, one under Information+Mo\-ti\-va\-tion exposure and the other under Information-only. Researchers followed the users during the walks at a suitable distance in order to take notes on specific behaviors (e.g., stopping to converse with SmartWalkCoach).
\rebadd{This in-the-wild observation was conducted as part of a field study approach, allowing the research team to systematically capture users’ natural interactions and behavior cues in context rather than relying solely on self-report measures~\cite{Stals2014Walkalong, Farrell2024FieldStudies}. Examples of such cues include spontaneous pauses to initiate interaction with SmartWalkCoach, body language indicating comprehension or hesitation, and real-time changes of pace or direction in response to system suggestions~\cite{Jorgensen2016Media}.} At the end of the experiment, participants shared their insights through mixed-method questionnaires and semi-structured interviews.

Our results show that the Information+Motivation condition dramatically improved users' subjective experience compared to the Information-only condition, with users registering significantly more positive feelings ($\beta=-0.806$, $p<.001$, $d=1.85$) and better user experience (the “User Experience” metric here is a composite, study-specific index like positive feelings rather than a generic UX measure). A thematic analysis of users' interviews identified `supportive companionship' and the `necessity of context-awareness' as the two key agent design tensions. The former was nurtured by timely, motivational speech, whereas transgressions of the latter by ill-timed interventions exposed the key design imperative of perceiving and responding to the user's current environment. These themes have direct design implications for how to optimize the expression/frequency/timing of accompanying agents during walking.

In this paper, we make the following contributions:
\begin{itemize}
    \item We introduce \emph{SmartWalkCoach}, a whole-system, end-to-end, intelligent walking companion that gives real-time, contextually aware support. It coordinates specific agents for route planning, in-walk coterie with dynamic, informational inspiration, and post-activity summary.
    \item We carry out a highly-controlled user study with a counterbalanced AB/BA crossover design to test our system in the field. With Linear Mixed Models and thematic analysis, we quantitatively and qualitatively demonstrate that contextually-informed motivation greatly increases user experience and positive affect compared to an information-only control. 
    \item We contribute towards HCI research by providing a design structure and empirical support for crafting effective mobile coaching systems, which show that combining relational tone and contextually aware timing has a key role to play in user engagement in mHealth systems. 
\end{itemize}

\section{Related Work}

\subsection{Conversational Agents for Physical Activity and Well-being}

Conversational systems aimed at changing health-related behavior -- including  physical activity, weight control, and eating habits -- have gained considerable attention in recent years~\cite{Oh2021SystematicReviewChatbotsPhysicalActivity}. At the mechanistic level, studies have incrementally established a ``motivation-goals-feedback'' design framework with explainable feedback and goal decomposition emphasis~\cite{Zhang2020AIChatbotModel}. At the inter-personal level, the consideration of relational styles has been found to improve user engagement and the system usability~\cite{Sage2025Companionship}. Moreover, research related to mobile health has established the importance of contextual alignment and timing sensitivity in supporting users' self-monitoring processes while minimizing interruptions~\cite{Chen2024ContextualNotifications}. Finally, in the wider context of goal-setting and planning, Abbas~\emph{et al.}~\cite{Abbas2025PITCH} revealed the ``wear-out effect'' of repetitive and predictable interactions with conversational agents. These results all collectively point to the conclusion that, in order to ensure their long-term effectiveness in daily real-world settings, conversational agents must rhythmically, materially, as well as pragmatically interact with user contexts. Our work thus places the Information+Motivation AI assistance within everyday walking accompaniment, extending work from previous guidance-based personalized walk-coaching systems~\cite{Raghuveer2022PACE}. 

\rebadd{Prior work has also investigated mobile health applications and personalized navigation practices to support walking activity among older adults. Felberbaum \emph{et al.} conducted a field study to derive design recommendations for mobile apps that encourage walking by addressing diverse walking variables, supporting long-term engagement, and enabling user control~\cite{Felberbaum2023MobileHealth}. Relatedly, Felberbaum’s CHI ’20 extended abstract demonstrated the value of integrating personalized navigation practice to enhance physical activity and dual-task walking performance in older adults~\cite{Felberbaum2020CHI}. These design insights highlight the importance of personalized, context-aware mobile support for walking, which aligns with our focus on context-rich, personalized walking companionship.}

\subsection{Tool-Using LLM Agents and Geospatial Copilots}

The tool-using intelligent agent paradigm, in which a language agent interacts with external tool APIs to map situational comprehension to executable actions, has drawn much interest in recent years~\cite{Xu2025LLMToolLearningSurvey}. Classic contributions to the field include \emph{Toolformer}, which first exhibited that large language models (LLMs) can acquire the ability to call external tools via self-supervised learning~\cite{Schick2023Toolformer}, and \emph{ReAct}, which combines reasoning and acting in an interleaved style~\cite{Yao2023ReAct}. Later systems, such as \emph{Gorilla} and \emph{ToolLLM}, have extended these functionalities further, allowing for reliable, scale-scalable connections to application programming interfaces (APIs) \cite{Patil2024Gorilla, Qin2024ToolLLM}

Within geospatial areas, this paradigm has been used to build specialized assistants. Systems like \emph{GeoAgent} utilize geocoding, POI search, and route calculation tools along with an LLM to normalize addresses and answer spatial queries~\cite{Huang2024GeoAgent}. Going beyond single-agent systems, there have been recent explorations of multi-agent architectures. For example, \emph{GeoLLM-Squad} outlines a multi-agent remote-sensing geospatial copilot that assigns subtasks to specialized agents~\cite{Lee2025GeoLLMSquad}. Again, \emph{MapAgent} outlines a geospatial-reasoning agent architecture that separates planning and execution explicitly and separates it dynamically using map-tool integration~\cite{Hasan2025MapAgent}.

Inspired by these developments, our system realizes a \textbf{lightweight event-driven and event-driven toolchain} across mobile devices. This design integrates modules for POI acquisition, route candidate generation, and dialogue synthesis. It generates a system that can sustain the low-latency, context-rich conversational interactions that occur in real-world walking.

\subsection{POI Recommendation and Route Planning Systems}

In recent work, Kim \emph{et al.} proposed a zero-shot learning approach for POI recommendations, which showed potential to provide customized suggestions without deep user history \rebadd{by extracting multi-aspect preferences from shallow data using an LLM~\cite{Kim2025ZeroPOIRec}.}
Our \emph{GeographyAgent} is conceptually related to this work, by combining collected user preferences as well as contextual reasoning to propose suitable routes, despite only having access to a shallow or non-existent user history.
Based on prior work in conversational route planning~\cite{MapGPT2025,ChatMap2023}, some systems have investigated natural-language-based user interfaces for navigation work. The GA-LLM framework~\cite{Liu2025GA-LLM} illustrates the capability of large language models to augment geographical thought by way of structured tool invocation, very much echoing our \emph{GeographyAgent}'s design approach. Likewise, adaptive navigation systems~\cite{Pacheco2022Systematic} have illustrated the potential benefit from dynamic route tailoring according to real-time contextual considerations.
Our work is novel in that we pay exclusive attention to walking companionship, where aesthetic value, pedestrian amenity, and ad-hoc discovery are given primacy over classical metrics of navigation such as travel time optimality. This is complemented by recent work on active transportation networks where well-being is given priority over efficiency~ \cite{Venkatraman2021RealTimeRouting}.

\subsection{Companionship and Motivational Systems}

Research on AI companions designed to promote physical activity has proceeded in multiple dimensions. The application of geofencing to trigger behavior interventions has been studied in previous work, where Shevchenko \emph{et al.}~\cite{Shevchenko2024Geofencing} showed location-based triggers to be effective for delivering context-specific notifications. This work furnishes the basis for the spatial splitting of interventions. Within the area of physical activity just-in-time interventions, Sporrel \emph{et al.}~\cite{Sporrel2022JITPrompts} showed the ``few but accurate'' rule for the delivery of prompts, highlighting the benefit of infrequent but accurately-crafted interventions over frequent but potentially disruptive messages. 
Our work here is also consistent with general models for the correctional aspects of behavior interventions during the execution of an overall plan, \rebrep{which stress the importance of timing for behavior interventions~\cite{Hardeman2019Systematic}}{which align with principles of JITAIs — a behavior change framework that emphasizes delivering tailored support at moments of need to maximize effectiveness~\cite{Hardeman2019Systematic}.}

Moreover, the relational design features of conversational chatbots have been well-examined within health settings. Bickmore's seminal work~\cite{Bickmore2010RelationalAgents} on the relationship features of relational agents emphasizes the need for empathic, sociable interactions. Later work~\cite{Sage2025Companionship} then examined how these principles extend specifically to physical activity coaching, where the most important dimensions are care, persistence, and exploration. Our work extends these foundations by using geofencing to achieve spatial segmentation of the route, by applying the ``few but accurate'' principle to the in-walk conversation, \rebrep{and
by extending the aforementioned relational design to our specific use-case of a walking companion}{where prompts are not sent at random or fixed intervals but are triggered only by contextual factors (e.g., user proximity to a goal or timing within the walk). This is consistent with research showing that well-timed, context-relevant prompts are more likely to support behavior change and engagement than frequent, poorly timed messages~\cite{Sporrel2022JITPrompts}.}

\section{Design and Implementation}

Our system, \emph{SmartWalkCoach}, is designed to provide an individualized walking experience through an end-to-end intelligent user interaction (IUI) workflow. It delivers real-time, context-aware AI companionship across the entire walking journey: from pre-walk planning and in-walk support to post-walk summarization~\cite{Abbas2025PITCH, Sage2025Companionship, Sporrel2022JITPrompts, Raghuveer2022PACE}.

The system architecture runs to a pipeline of a sequence of three specialized, non-communicating functional agents--the \emph{GeographyAgent}, \emph{AccompanyAgent}, and \emph{SummaryAgent}\rebadd{, together with a light-weight bridging agent}. Each functional agent runs in its own stage of the process and does not call or message the other agents directly. Rather, they communicate indirectly by writing to and reading from a shared, structured state object (e.g., the selected route, user preference, walk statistics). \rebrep{By this design, it achieves modularity and strict separation of concerns}{Strictly separating the three agents creates clear security boundaries that limit each agent’s access and capabilities, reduce shared trust assumptions, and make it easier to audit and control interactions—thereby lowering the risk that malicious or adversarial behavior in one agent can compromise the system as a whole~\cite{Del2025Architecting}}.

Alongside the three functional agents, our design includes a \emph{user-oriented lightweight bridging agent}. This agent serves as the sole communication interface between the user and the functional agents. It has the primary roles of receiving all user input, finding out the current context of the interaction, sending commands to the functional agents, and receiving text responses. \rebadd{This is achieved through token-based orchestration of the three functional agents' output (see Appendix.}

In summary, the functional agents (\emph{SummaryAgent}, \emph{AccompanyAgent}, \emph{GeographyAgent}) are the information providers and decision-makers, which run over shared state. The bridging agent is the coordinator which handles the workflow and has a specialty in user-centric communication.
All four agents are instantiated based on OpenAI's GPT-4o API, and are prompted to perform their respective roles. This paradigm follows recent language-driven tool invocation paradigms \cite{Schick2023Toolformer, Yao2023ReAct, Patil2024Gorilla, Qin2024ToolLLM}.

\subsection{System Overview and Process Orchestration}

Drawing inspiration from recent agentic conversational planning architectures~\cite{Abbas2025PITCH}, our mobile client uses a unified in-app orchestration with event-based delivery to minimize user work but maintain precise control over what to show and when to show it. \rebadd{Event-based delivery reduces user workload by avoiding repetitive or irrelevant prompts and facilitates precision because events encode meaningful user or context transitions that directly trigger agent actions~\cite{Mehrotra2015IntelligentNotifications}.}
 A lightweight runtime system continuously collects and collates context signals (location/geofence, cadence, mileage, etc.). On occasion of decisions points, the system invokes \emph{AccompanyAgent} to construct up-to-date status reports or recommendations. The results are then pushed over event channels to the front-end for display and logging, allowing auditability and post-hoc analysis. 
 
 This design employs JITAI principles for timing and tailoring~\cite{Hardeman2019Systematic} and can be assessed through micro-randomised trials for component optimisation~\cite{Klasnja2015MRT}. To avoid excessive prompting and encourage user acceptance, the scheduler incorporates principles from geofencing and notification receptivity research~\cite{Shevchenko2024Geofencing,Pielot2014Notifications,Mehrotra2016Receptivity}, while the response tone is guided by relational-agent work to ensure continued engagement and user satisfaction~\cite{Bickmore2010RelationalAgents}. For heavier tasks (e.g., summarizing or planning, tool invocation), we make use of straightforward multi-agent designs with role specialisation and reflective turns (see e.g.,~\cite{Wu2024AutoGen,CAMEL2023}), but retain all the orchestration on the client-side. This ensures deterministic timing with minimal user inconvenience.

\subsection{GeographyAgent}

\emph{GeographyAgent} is tasked with devising suitable walking endpoints and waypoints through consideration of user needs and preferences, together with existing POI choices. Its only role is to determine a specifically curated JSON list of POIs~\cite{Singh2024GeoLLMEngine}, with the actual route planning -- path discovery, distance calculation, navigation cues -- are outsourced to commercial map service professional tools through standard APIs. This strict separation allows the LLM to focus on semantic interpretation and reasoning tasks based on user preferences, while leaving the heavy lifting of the complex geospatial calculations to the specialized external tools~\cite{Singh2024GeoLLMEngine,Huang2024GeoAgent}.

Our design therefore applies a strict separation of concerns, following the same process as similar tool-assisted LLM paradigms in previous work~\cite{Schick2023Toolformer,Yao2023ReAct}. A light-weight language-action intermediary manages the interlingual translation from natural language to tool calls. This intermediary tokenizes user input to structured action tokens and intent indicators (e.g., \texttt{poi.nearby}, \texttt{route.alternative}) and forwards them to suitable geographical tools~\cite{Patil2024Gorilla,Qin2024ToolLLM}. If POI querying is required, our system employs the following hierarchical strategy: database-first querying, on-the-fly API probing, minimal gap filling. \emph{GeographyAgent}'s principal role remains preference-conscious POI selection from preselected candidates. This is accomplished through logical reasoning where the model evaluates and ranks POI names according to their accessibility and attractiveness, together with \rebrep{user-profile-based personalized preference weights}{user preferences.} This approach ensures that the agent's focus remains solely on selecting suitable POI names, without the overhead of geographical computations~\cite{Kim2025ZeroPOIRec,Liu2025GA-LLM}. 
In order to accommodate varied user scenarios, we deploy a cross-situational reasoning process. \rebrep{If users submit sketchy intentions, the system provides both the user profiles and attainable POI candidates into the LLM, so personalized suggestions can be generated even when the input is under-specified.}{The agent will firstly judge whether the user has a clearly directed destination. If yes, the system will head towards the branch that has been filtered from the candidate POI list. Otherwise, if users submit unclear intentions, the system provides both the user profiles and attainable POI candidates into the LLM for logical reasoning, so personalized and reasonable suggestions can be generated.} This versatility results from recent development advances for LLM-based recommendation agents that sustain well-consistent reasoning with multiple modalities of interaction~\cite{Xiao2025MMAgentRec,Huang2025AgenticRecommender,Wu2025Tracemate}.

\begin{figure}[!htbp]
  \centering
  \makebox[0pt][c]{%
    \includegraphics[width=1.3\linewidth,
      trim=1.2cm 0 1.2cm 0, clip]{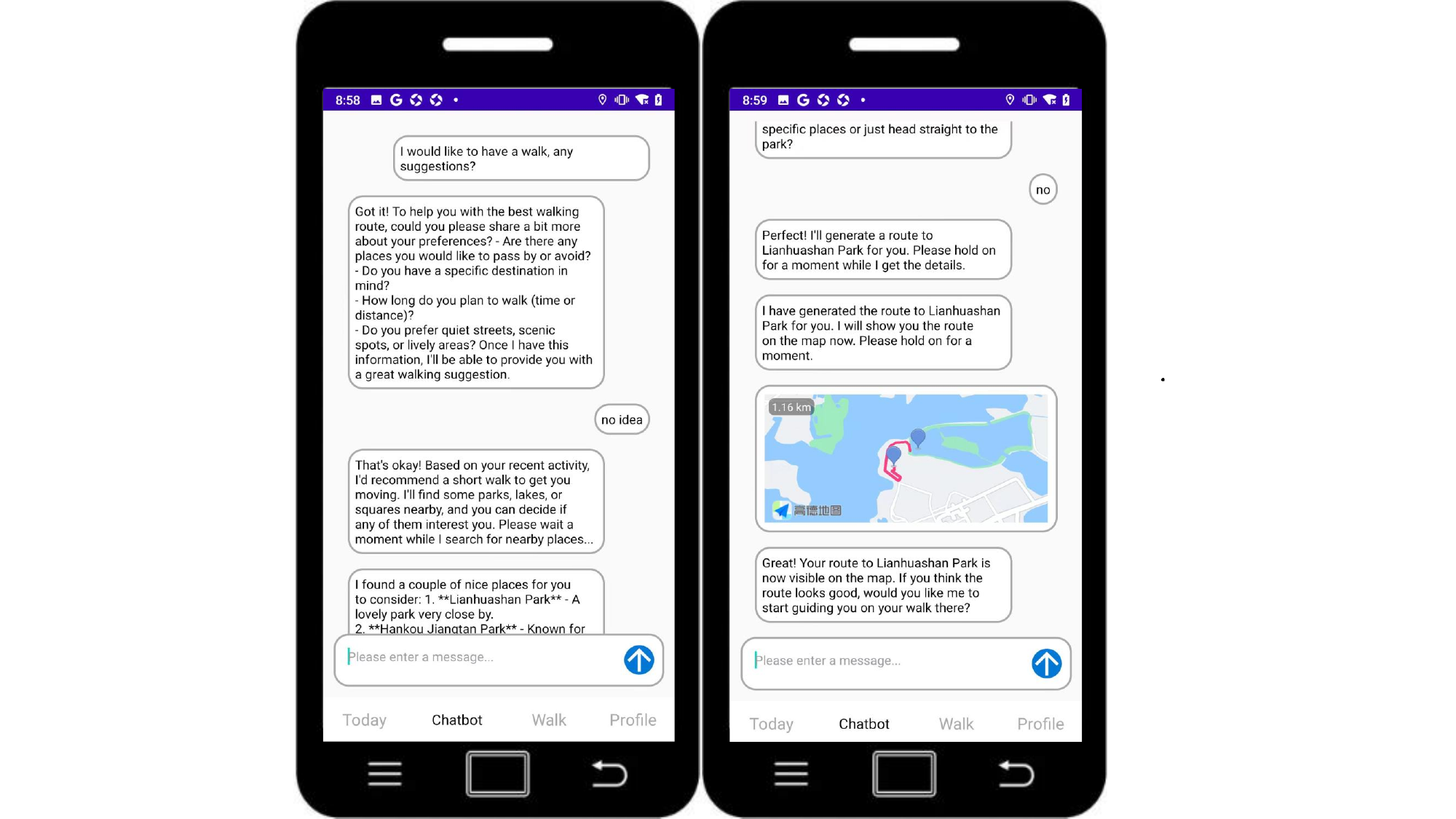}}
  \Description{The image shows an interaction between the user and SmartWalkCoach during the pre-walk route planning stage.}
  \caption{Screenshots of the interaction in the pre-walk route planning stage.}
  \label{fig:planning}
\end{figure}

Once \emph{GeographyAgent} has generated a suitable shortlist of appropriate POIs, the language-action intermediary constructs an exact request to routing APIs, which actually complete the pathfinding action, and return routable paths. This strategy for isolating LLM reasoning from geographical calculation is consistent with common designs for secure tool-augmented language models~\cite{Schick2023Toolformer,Yao2023ReAct,Patil2024Gorilla}.
Figure~\ref{fig:planning} shows an example of a posssible interaction between a user and the system during the pre-walk route planning stage.

\rebdel{This design architecture provides an answer to our RQ1, by showing how conversational interfaces decrease cognitive load through role-specialization for agents.} By narrowing the role of \emph{GeographyAgent} to handling POI selection alongside complementary external planning service implementations, the overall system allows for intuitive interactions with users, where they provide their preferences using natural language rather than navigating complex menus or map controls.

\subsection{AccompanyAgent}\label{subsec:AccompanyAgent}

\emph{AccompanyAgent} is tasked with walking companionship and user motivation. It holds the user's proximate input and current context within the planned walking route, is given recent location and progress summaries through a tick-based system, and pushes new POI annotations or phase reports back to the conversation and UI through callbacks. As such, the agent's primary task is to acquire contextually relevant information such as users’ walking progress and speed, or nearby preferred POIs. We chose this approach because empirical evidence suggests that interventions combining informative feedback with encouragement are more effective at promoting exercise and physical activity than undifferentiated encouragement alone~\cite{Ghantasala2023}.

The system adopts a distance-segmented geofencing mechanism for timely, contextually appropriate prompting: the walking path is segmented in real-time, and on entering each new segment, informative or motivational prompting is triggered. \rebadd{The triggered segment length is determined by the user’s preferred prompting frequency (i.e., how often they are willing to be disturbed during walking). If the user prefers fewer messages/alerts, we use longer segments to reduce interruptions; conversely, we shorten segments when the user opts for more frequent prompts.} This virtual geofencing approach extends similar previous work~ \cite{Shevchenko2024Geofencing,Tobin2023Geo,Sasaki2024Data}.

AccompanyAgent realizes a real-time rhythm detection function for awareness of \rebrep{driver}{walker} fatigue. It continuously computes the average walking pace of the user based on a sliding window. If there has been a sharp decrease in pace between the current window and a reference from a prior stage, a dynamic intervention will be initiated. \rebrep{It will comprise a real-time status update along with reassurance or encouragement.}{We incorporate positive reinforcement messages and real-time progress updates as part of our adaptive intervention language, intended to motivate continued walking and support sustained engagement~\cite{Wendy2019JITAI}.} To be responsive, the complete backend pipeline is designed to run with low-latency processing and lightweight, clear structured output utilizing a shared token data contract. By doing this, the front-end interaction level has parsable complexity reduced. The tone and timing of these motivational questions follow the few but correct rules of real-time interventions \cite{Sporrel2022JITPrompts} and are intended to be relational-chatbot-proof following the care—persistence—exploration guidelines to enable a relational sense of social connection \cite{Sage2025Companionship}. 

Contrary to frequent navigation maps with high-density annotations, we adopt a minimalist map view approach during the accompaniment stage. This means that we obliterate most POI layers and heavy-duty controls, and only keep the vital cues pertaining to the present task (directions, paths, and so on). The environment information and search loads are managed by the agent in the background, and incrementally divulged as short prompts or annotations. As such, our design follows the principle to ``delegate complexity to the agent, leave essentials to the user'', resulting in reduced cognitive load. In particular, this reduces the amount of \rebrep{eye-switching necessary during the walk}{ attention needed on map}, so that the user can pay more attention to their walking activity \rebadd{or other phone tasks (e.g., chatting, scrolling)}. \emph{AccompanyAgent} only intervenes when the virtual geofence or a rhythm event are triggered, in which case it presents the user with a small set of high-value information such as one or two nearby candidate rest points, or some suggested small correction to the user's route. The low-frequency but timely nature of these interventions allows the user to focus on the big picture instead of them being overloaded with unimportant information.

\begin{figure}[!htbp]
  \centering
  \makebox[0pt][c]{%
    \includegraphics[width=1.3\linewidth, trim=1.2cm 0 1.2cm 0, clip]{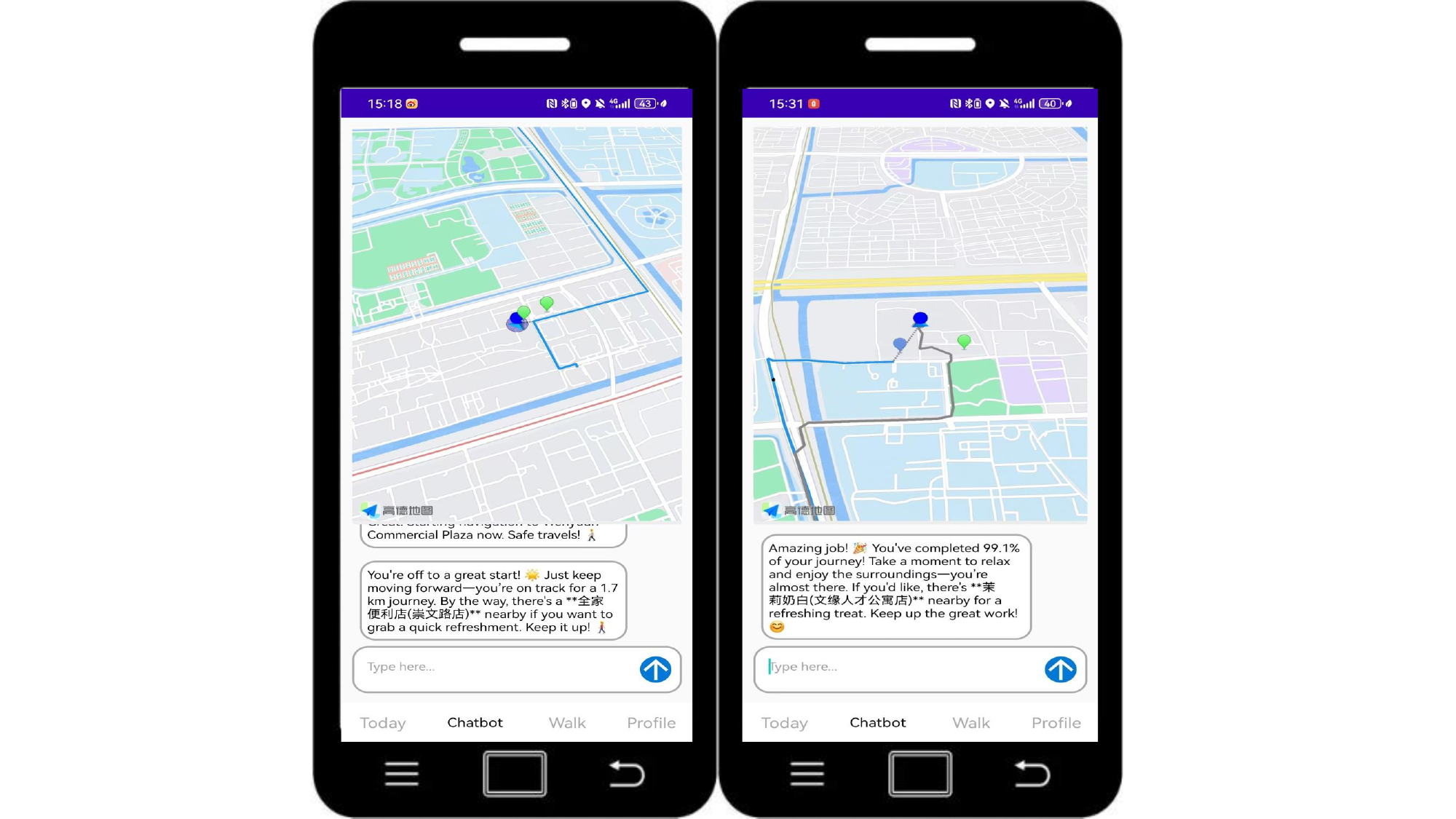}
  }
  \Description{The image shows an interaction between the user and SmartWalkCoach during the walk.}
  \caption{Screenshots of an interaction between the user and SmartWalkCoach during the walk.}
  \smallskip 
  \parbox{0.9\linewidth}{\small\itshape 
    The Chinese characters in the picture all represent POI names and the green markers represent POI locations:
    the one on the left represents a supermarket, and the one on the right represents a milk tea shop.
  }
  \label{fig:stage2_interaction_detailed}
\end{figure}

Regarding the POI search, \emph{AccompanyAgent} uses the same fundamental methodology as \emph{GeographyAgent}, but runs on the spatially restricted scope set by the minimum geofence distance. The agent is given the narrow-down POI table over this limited region, and uses contextual inference tied to user historical likes, present requirements, and contextual hints to narrow down the POIs to the most suitable ones. It then formats the chosen POI data to specific output templates. These are simultaneously processed by the system for map display, and passed to the user-facing lightweight bridge agent for suitable presentation to the user through conversational means. Figure~\ref{fig:stage2_interaction_detailed} shows an example of a possible interaction between the user and SmartWalkCoach during the walk \rebadd{(see more examples in the appendix)}.

\subsection{SummaryAgent}

When a walk nears completion, the orchestration layer emits a structured payload (mileage, duration, goal attainment) and invokes a minimal \emph{SummaryAgent} to:
\begin{description}
\item[(i)] \textbf{Generate a brief walk summary} that emphasizes user accomplishments and offers helpful feedback. This feature conforms to evidence-backed behavior change techniques in health behavior change, as systematic feedback production has been found to have a substantial impact on enhancing self-monitoring activity and exercise behavior \cite{Krukowski2024_FeedbackMetaAnalysis, Davey2015_ImprovingDesignReportingBehaviourChange}. By presenting prompt, achievement-oriented feedback, the summary reminds the user of their advancement and encourages good behavior patterns.

\item[(ii)] \textbf{Suggest a specific follow-up step based on if--then action planning} to \rebrep{address the intention-behavior gap}{ encourage post-walk reflection and consolidation}. This strategy builds on the well-established evidence supporting implementation intentions which commit to when, where, and how a behavior will be initiated. Recent systematic reviews verify that action planning interventions yield consistent improvements in a range of health behaviors by establishing strong mental connections between situational cues and intention-driven responses \cite{Hagger2014_ImplementationIntentionReview, Bailey2017_GoalSettingActionPlanning}.

\item[(iii)] \textbf{Optionally externalize key records to social outlets} to leverage social accountability mechanisms. This feature has its theoretical base in behavior change technique design frameworks that value public commitment and social support to maintain behavior \cite{Davey2015_ImprovingDesignReportingBehaviourChange}. By allowing users to externally post their accomplishments voluntarily, the system provides social reinforcement while being sensitive to individual privacy needs, thereby promoting continued participation over the long term based on external validation and social support.
\end{description}

The \emph{SummaryAgent} thereby implements several evidence-based behavior change techniques---feedback, planning, and social support---within a lightweight, computer-automated process that imposes minimal user effort while optimizing the building of walking successes into sustained habits.

We include this agent chiefly to close the loop and keep the pipeline end-to-end: logging, concise reflection, and a concrete next action. This choice is supported by recent evidence that implementation intentions (if–then plans) yield reliable improvements across cognitive, affective, and behavioral outcomes, with larger effects when plans specify time and place cues and are briefly rehearsed~\cite{Sheeran2025WhenHow}. In mobile physical-activity contexts, Vetrovsky \emph{et al.}~\cite{Vetrovsky2022SelfMonPlus} showed that adding structured goals and plan components to self-monitored activities produces additional and sustained health gains over self-monitoring. We also keep the summary short and tie it to a specific next cue to avoid reflection-only burden, and make social posting optional because social incentives can be context-dependent or reduce over time~\cite{Agarwal2021Gamification,Patel2021Gamification}.

\section{Methods}

\subsection{Experimental Design}

We employed a counterbalanced AB/BA crossover design to compare the Information-only versus Information+Motivation conditions. Here, Information+Motivation refers to the condition where the agent augmented basic navigational guidance with motivational and relational strategies. These included using a companionable or coaching tone, providing positive feedback on progress, addressing the user by name, recalling stated preferences. In contrast, the Information-only condition delivered strictly factual navigational content without any encouraging or relational phrasing, consistent with the operational definitions used in recent studies of relational agents for physical activity \cite{Sage2025Companionship, Raghuveer2022PACE}. The counterbalancing introduced minimizes sequence and carryover bias, and this overall design improves efficiency by utilizing a within-subjects approach, where participants serve as their own controls across the two conditions~\cite{JonesKenward2014,Capili2024Crossover}. We also probed period×treatment interactions and assessed potential behavioral carryover as suggested in recent guidance~\cite{Shi2024Carryover}. 

Because this study was conducted in-the-wild (i.e. outside laboratory constraints, in participants’ natural walking environments) and adopted a walk-along style of accompaniment, we expect ecological validity but accept constraints on sample size \cite{Harries2013_WalkingInTheWild, Amaya2022_WalkAlongReview}. Twelve participants completed both conditions (N = 12), yielding 24 total observations \rebadd{(see Table~\ref{tab:demographics})}. \rebadd{Ethical approval for this study was obtained from the authors' institutional Ethics Review Panel.}
\begin{table}[!htbp]
  \centering
  \rebadd{%
  \caption{Demographic Characteristics of Study Participants}
  \begin{tabular}{l c}
    \toprule
    \textbf{Characteristic} & \textbf{Value} \\
    \midrule
    Total participants (N) & 12 \\
    Gender (Male / Female) & 5 / 7 \\
    Age (Mean ± SD, range) & 24.17 ± 10.87 (17–47) \\
    Walking Habit (Frequent / Occasional) & 9 / 3 \\
    Tech Familiarity (High / Medium / Low) & 10 / 2 / 0 \\
    \bottomrule
  \end{tabular}
  \label{tab:demographics}
  }
\end{table}

\subsection{Data collection and analysis}

We selected validated short instruments where possible and defined two lightweight, study-specific constructs to capture design goals (see Table~\ref{tab:survey_constructs}). We measured baseline affect before the experiment using the 10-item I-PANAS-SF~\cite{Thompson2007IPANASSF}. To contextualize these quantitative measures, researchers conducted walk-along observations, with recorded behavioral data used solely for verification. 
The measurements on post-segment motivation and enjoyment use items from the Intrinsic Motivation Inventory (IMI)~\cite{SDT_IMI_Instrument}, focusing on interest/enjoyment and perceived choice facets, with minor wording adaptations for in-walk use. User engagement is measured with the User Engagement Scale short form (UES-SF)~\cite{OBrien2018UES}. In addition, we operationalized \emph{Alignment and Sense of Control}, including elements such as perceived fit, timing, and how easy the system is to override or customize, and \emph{Companionship and Relational Bond}, including perceived caring tone and supportive presence, as brief study constructs. \rebadd{These were chosen to sustain walking when users feel hesitation or fatigue. Indeed, prior HCI work on coaching-style conversational systems and relational agents suggests that supportive, autonomy-respecting language (e.g., encouragement, acknowledgment, and non-pressuring suggestions) can increase continued engagement in activities~\cite{Sage2025Companionship}.} Survey items for these two constructs \rebadd{(see Appendix)} were informed by contemporary HCI work emphasizing explicit control and alignment in tool-augmented conversational systems~\cite{ACM2024Alignment} and by relational-agent coaching evidence in mobile health~\cite{Sage2025Companionship}.

\begin{table*}[!htbp]
\caption{Categories, dimensions, and descriptions of quantitative data collected in the study surveys.}
\label{tab:survey_constructs}
\centering
\small
\begin{threeparttable}
\begin{tabular}{p{0.28\linewidth} p{0.22\linewidth} p{0.42\linewidth}}
\toprule
\textbf{Construct} & \textbf{Type} & \textbf{Description} \\
\midrule
Baseline (I-PANAS-SF: PA \& NA)\tnote{a} & 5-point Likert (10 items; pre-walk) & Positive and negative affect before walking; used as a baseline covariate. \\
Motivation/Enjoyment (IMI subscale)\tnote{b} & 7-point Likert (5–6 items; post-segment A/B) & Interest/enjoyment and perceived choice during the walking–system interaction. \\
User Engagement (UES-SF)\tnote{c} & 7-point Likert (6 items; post-segment A/B) & Focused attention, clear feedback, low effort, and willingness to persist. \\
Alignment \& Sense of Control\tnote{d} & 7-point Likert (6 items; post-segment A/B) & Perceived fit, timely prompts, and easy adjustment or override of frequency and timing. \\
Companionship / Relational Bond\tnote{e} & 7-point Likert (6 items; post-segment A/B) & Perceived supportive companion, sincerity of encouragement, and willingness to reuse or recommend. \\
\bottomrule
\end{tabular}
\begin{tablenotes}\footnotesize
\item[a] I-PANAS-SF short form \cite{Thompson2007IPANASSF}.
\item[b] Intrinsic Motivation Inventory (interest/enjoyment, perceived choice) \cite{SDT_IMI_Instrument}.
\item[c] User Engagement Scale short form (UES-SF) \cite{OBrien2018UES}.
\item[d] Study-specific items informed by explicit control/alignment in tool-augmented conversational systems \cite{ACM2024Alignment}.
\item[e] Study-specific items informed by relational-agent coaching evidence in mobile health \cite{Sage2025Companionship}.
\end{tablenotes}
\end{threeparttable}
\end{table*}

To complement the quantitative measures, we ran a brief semi-structured interview with each participant after they had used the app for 1–2 hours in the experiment, following recent HCI practice and mobile relational-chatbot studies~\cite{ACM2024Alignment,Sage2025Companionship}. These interviews lasted 10–15 minutes and were audio-recorded with verbal consent. We encouraged concrete situational detail (where/when/what the prompt said, and how they responded). Notes focused on wording, trigger timing and channel, perceived fit or burden, and causes of differences. Our evaluation also adhered to standard qualitative research guidelines~\cite{Crabtree2025EvalQualHCI,Nowell2017ThematicAnalysis}, giving tangible direction for the maximization of the role of the accompanying agent by the means of context.

\subsection{Composite Variable Construction}

In order to provide strong measurements, we built composite variables informed by well-established theoretical frameworks. \rebadd{Specifically, we grouped items into two subsets: \textbf{positive feelings} (affective/relational appraisal, e.g., enjoyment, perceived value, and supportive presence) and \textbf{usage experience} (pragmatic/functional interaction quality, e.g., attention, clarity, effort, and perceived control). This split follows the common psychometric practice of aggregating conceptually coherent indicators into higher-level composites for construct validity, and aligns with HCI work that treats perceived alignment/control as a distinct functional dimension of interactive experience \cite{Clark2019,Sage2025Companionship}.} 

The \textbf{positive feelings} variable combined six psychometric indicators from the post-session questionnaires: enjoyment and pleasant attitude from the Intrinsic Motivation Inventory (IMI), worthwhile-ness from the User Engagement Scale (UES), connection and sincerity from the Companionship/Relational Feeling scale \cite{Sage2025Companionship}, and continuation intention from IMI. Meanwhile, the \textbf{usage experience} variable was constructed from six complementary indicators focusing on the pragmatic and functional aspects of the interaction: engagement (IMI), focused attention, willingness to continue, ease of use, and clarity (UES), along with perceived customization and control from the Alignment scale. The multi-scale method, guided by concepts from alignment in interactive systems \cite{ACM2024Alignment}, is compatible with current psychometric practices which focus on construct validity through composite scoring \cite{Clark2019}. Before conducting the main hypothesis tests, we assessed the internal reliability of our composite measures. The positive feelings dimension demonstrated excellent reliability (Cronbach's $\alpha = 0.916$), with strong average inter-item correlations ($r = 0.650$). Likewise, the usage experience dimension showed good reliability ($\alpha = 0.881$), supporting its adoption as an instrument-based measure for mechanistic functional experience enhancements.

\subsection{Statistical Analysis: Linear Mixed Models (LMM)}

Our statistical modeling employed linear mixed models (LMM) estimated by restricted maximum likelihood (REML), excluding any bias from the variance component estimates for repeated measures designs~\cite{Bates2015}. The model specification was treatment condition (Information-only vs.\ Information+Motivation), sequence order (AB vs.\ BA), and period (A vs.\ B), as fixed effects, with participant-level random intercepts to permit individual variation. One notable methodological consideration for crossover designs is carryover, with the effects from the first treatment still present during the outcome during the treatment. We adjusted for this by including the treatment × sequence interaction term from our first model specification. As is standard practice for AB/BA designs~\cite{Senn2002}, the absence of any significant interaction would indicate the absence of informative carryover effects, allowing direct treatment effects to be validly interpreted.

\section{Results and Discussion}

\subsection{Validation of Models and Carryover Effects}

The preliminary model diagnostics showed no substantial carryover effects, as the treatment × sequence interaction was not significant ($\beta = -0.444, p = 0.137$). This result is good confirmation of our crossover design's key assumption that the treatments are independent of each other. The result is also good confirmation of the interpretation of direct period effects. That there are no carryover effects is methodologically desirable, since the apparent treatment effects are not contaminated by enduring effects from previous interventions~\cite{Senn2002}.

\subsection{Effects on Positive Emotions}

The linear mixed model test had a statistically significant and large main effect of the Information+Motivation (Info-Motive) manipulation on positive feelings. As presented in Table~\ref{tab:lmm_positive_feelings}, the Info-Motive group had significantly larger positive feelings compared to the Info-Only group ($\beta = -0.806, p < 0.001$), representing an average increase by 1.028 points on the composite measure. The effect size was assessed by Cohen's d and equaled 1.845, reflecting by traditional criteria a large effect size~\cite{Cohen1988}.

Further inspection of Table~\ref{tab:lmm_positive_feelings} finds the following key patterns. The highly significant intercept ($\beta = 5.417, p < 0.001$) suggests the presence of strong positive feelings during the Info-Motive condition. The negative coefficient for Info-Only suggests a strong drop-off in positive feelings when the motivational factors were stripped away. We observed a significant sequence effect (β = 0.889, p < 0.001), indicating that participants in the BA sequence tended to report higher average positive feelings across both periods. The treatment × sequence interaction did not reach statistical significance (p = 0.137), which provides some tentative evidence against a strong carry-over effect, under our model assumptions. 

\begin{table*}[!t]
\caption{Linear Mixed Model Results for Positive Feelings}
\label{tab:lmm_positive_feelings}
\centering
\small
\begin{threeparttable}
\begin{tabular}{lccccc}
\hline
\textbf{Parameter} & \textbf{Coefficient} & \textbf{SE} & \textbf{z-value} & \textbf{p-value} & \textbf{95\% CI} \\
\hline
\multicolumn{6}{l}{\textbf{Fixed Effects}} \\
\hline
Intercept & 5.417*** & 0.179 & 30.197 & <0.001 & [5.065, 5.768] \\
Info-Only & -0.806*** & 0.211 & -3.814 & <0.001 & [-1.219, -0.392] \\
Sequence (BA) & 0.889*** & 0.254 & 3.504 & <0.001 & [0.392, 1.386] \\
Treatment × Sequence & -0.444 & 0.299 & -1.488 & 0.137 & [-1.030, 0.141] \\
Group Var & 0.443 & 0.596 & 0.743 & 0.458 & [-0.726, 1.612] \\
\hline
\multicolumn{6}{l}{\textbf{Random Effects}} \\
\hline
Intercept Variance & 0.059 & - & - & - & - \\
Residual Variance & 0.134 & - & - & - & - \\
\hline
\end{tabular}
\begin{tablenotes}\footnotesize
\item \textit{Note:} *$p < 0.05$, **$p < 0.01$, ***$p < 0.001$. 
N = 24 observations from 12 participants. 
Model estimated using REML. Log-Likelihood = -15.020.
\end{tablenotes}
\end{threeparttable}
\end{table*}

\subsection{Mechanistic Evidence from User Experience}

The Info-Motive intervention improved user experience significantly ($\beta = -0.583, p = 0.007$), showing an extremely large effect sizes measure (Cohen's d = 1.464). An inspection of Table~\ref{tab:lmm_user_experience} provides the following important patterns: the strong negative Info-Only coefficient describes decreased user experience without motivational factors, the very strong intercept ($\beta = 5.278, p < 0.001$) describes the high level of baseline satisfaction. The very strong sequence effect ($\beta = 0.944, p < 0.001$) establishes order effects, the non-significance of the interaction term ($p = 0.149$) establishes the crossover design.

\begin{table*}[!htbp]
\caption{Linear Mixed Model Results for User Experience}
\label{tab:lmm_user_experience}
\centering
\small
\begin{threeparttable}
\begin{tabular}{lccccc}
\hline
\textbf{Parameter} & \textbf{Coefficient} & \textbf{SE} & \textbf{z-value} & \textbf{p-value} & \textbf{95\% CI} \\
\hline
\multicolumn{6}{l}{\textbf{Fixed Effects}} \\
\hline
Intercept & 5.278*** & 0.164 & 32.162 & <0.001 & [4.956, 5.599] \\
Info-Only & -0.583** & 0.218 & -2.680 & 0.007 & [-1.010, -0.157] \\
Sequence (BA) & 0.944*** & 0.232 & 4.070 & <0.001 & [0.490, 1.399] \\
Treatment × Sequence & -0.444 & 0.308 & -1.444 & 0.149 & [-1.048, 0.159] \\
Group Var & 0.137 & 0.403 & 0.340 & 0.734 & [-0.653, 0.926] \\
\hline
\multicolumn{6}{l}{\textbf{Random Effects}} \\
\hline
Intercept Variance & 0.019 & - & - & - & - \\
Residual Variance & 0.142 & - & - & - & - \\
\hline
\end{tabular}
\begin{tablenotes}\footnotesize
\item \textit{Note:} *$p < 0.05$, **$p < 0.01$, ***$p < 0.001$. 
N = 24 observations from 12 participants. 
Model estimated using REML. Log-Likelihood = -13.661.
\end{tablenotes}
\end{threeparttable}
\end{table*}

The random effects pattern exhibits reasonable individual variation (intercept variance = 0.019) but notable within-person variation (residual variance = 0.142), implying user experience was governed by situational aspects rather than stable characteristics. These results suggest motivational aspects significantly supplemented functional experience by affective gain.

\subsection{Qualitative Insights from Post-Study Interviews through Thematic Analysis}

Our thematic analysis of 12 semi-structured interviews provided rich insights into user experiences with prompt delivery, directly addressing RQ3 on optimizing agent expression, frequency, and timing. Interviews followed a standardized protocol focusing on timing, frequency, expression, and companion relationships, with transcripts analyzed using established thematic analysis methods \cite{Braun2006Using,Nowell2017ThematicAnalysis}. \rebadd{We conducted a hybrid deductive--inductive thematic analysis: transcripts were first coded using the four interview themes as an initial template (comparing, alignment/control, relational experience, and critical incidents), and sub-themes were then refined iteratively from the data within each template category~\cite{Braun2006Using}.
} 

\rebadd{Table~\ref{tab:thematic_analysis_results} shows the results of the thematic analysis.} A central finding concerned \textbf{expression modality and relational tone}: participants consistently preferred motivational language that fostered partnership over transactional interaction. 
As one user reported, ``The positive words made me feel like [I was interacting with] a companion instead of a navigator tool''. This affectionate tone was directly contrasted with the Info-Only tone, which was described as ``cold'' and ``impersonal''.

In terms of \textbf{timeliness}, respondents all agreed on contextual appropriateness. Out-of-place interventions during high-cognitive-load conditions such as street crossing dramatically lessened the experience. A one respondent noted, ``While attempting to cross the street, the last thing you want is notification of my pace''. This underscores the paramount importance for context-conscious mechanisms for timely behavior that take the user's safety first. The study also considered \textbf{ideal frequency patterns}. Participants preferred infrequent but insightful interactions, with one stating that ``A few well-placed prompts were much more effective than incessant chatter''. This indicates that frequency must coincide with important journey milestones instead of set intervals.

Users also wished for \textbf{fine-grained control over agent behavior}, seeking the ability to personalize frequency, tone, and timing. The idea repeats the emphasis on end-user agency over interaction management.
In addition, users indicated a clear preference for \textbf{multi-modal interaction}, with repeated demands for voice-associated modes of visual prompts. For example, one participant stated: ``If I could hear the prompts instead of glance at the phone, it would be safer and convenient''.
Finally, our interviews showed that \textbf{individualization and dynamism} of the app's responses were key to continued interaction with users. Users signaled their desire for systems to learn from their preferences and likes, and learn prompting tactics from them.

\begin{table*}[t]
\caption{Thematic Analysis of User Requirements for Optimizing Walking Companion Agents (RQ3)}
\label{tab:thematic_analysis_results}
\footnotesize
\begin{tabularx}{\textwidth}{p{2.2cm} p{2.3cm} >{\raggedright\arraybackslash}X p{3.2cm}}
\hline
\textbf{Theme} & \textbf{Sub-Theme} & \textbf{Representative User Voices} & \textbf{Design Implications \& Recommendations} \\
\hline

\multirow{4}{*}{\textbf{\parbox{2.2cm}{1.Expression Modality: From Instrumental to Relational}}} 
& \textbf{1.1 Humanized \& Encouraging Communication} 
& \begin{itemize}[leftmargin=*,nosep,topsep=0pt,itemsep=0pt]
    \item ``The second stage's prompts gave me some words of encouragement... It was more emotional and motivating.'' (P2)
    \item ``The first version felt more like a real person... The return trip mode was like a machine, just reporting data, cold and uninteresting.'' (P7)
    \item ``It felt like a friend, speaking in an interesting way and encouraging me.'' (P8)
  \end{itemize}
& \begin{itemize}[leftmargin=*,nosep,topsep=0pt,itemsep=0pt]
    \item \textbf{Agent persona should be relational, not purely instrumental.}
    \item Avoid repetitive, mechanical progress updates.
    \item Use varied, empathetic, and motivating language.
  \end{itemize} \\

\cline{2-4}
& \textbf{1.2 Supportive Autonomy vs. Perceived Control} 
& \begin{itemize}[leftmargin=*,nosep,topsep=0pt,itemsep=0pt]
    \item ``The route was set too rigidly... I had to follow it strictly. I prefer designs that allow for self-exploration.'' (P1, P4)
    \item ``It felt like being managed or urged.'' (P4)
    \item ``Ultimately, I decide how to walk; it's more like a talkative navigation.'' (P10)
  \end{itemize}
& \begin{itemize}[leftmargin=*,nosep,topsep=0pt,itemsep=0pt]
    \item \textbf{Re-frame directives as suggestions.}
    \item Replace commands with inquisitive prompts.
    \item Support minor deviations from planned routes.
  \end{itemize} \\

\hline

\multirow{4}{*}{\textbf{\parbox{2.2cm}{2.Timing: Context-Aware Intervention}}} 
& \textbf{2.1 Avoiding High Cognitive Load Situations} 
& \begin{itemize}[leftmargin=*,nosep,topsep=0pt,itemsep=0pt]
    \item ``At crossroads or places with heavy traffic, it's inappropriate... I have no time to look at the screen.'' (P1, P2, P3)
    \item ``In noisy environments or when crossing the street, I would ignore the prompts.'' (P3, P7, P8)
    \item ``When I'm chatting with a friend, a notification can be disruptive.'' (P10)
  \end{itemize}
& \begin{itemize}[leftmargin=*,nosep,topsep=0pt,itemsep=0pt]
    \item \textbf{Implement context-aware suppression of notifications.}
    \item Integrate sensors (GPS, microphone) to detect high-load contexts.
    \item Minimize alerts in safety-critical situations.
  \end{itemize} \\

\cline{2-4}
& \textbf{2.2 Proactive Support at Moments of Need} 
& \begin{itemize}[leftmargin=*,nosep,topsep=0pt,itemsep=0pt]
    \item ``When I was walking up a slope and getting tired, it said, 'You've done a great job,' which gave me energy.'' (P8)
    \item ``When I reached 3/4 of the way and was about to give up, its encouragement helped me persist.'' (P3, P7, P12)
    \item ``It could detect my pace slowed down and asked if I was tired. The timing was very accurate.'' (P2, P5)
  \end{itemize}
& \begin{itemize}[leftmargin=*,nosep,topsep=0pt,itemsep=0pt]
    \item \textbf{Leverage behavioral and physiological cues for timely support.}
    \item Use triggers (step frequency, heart rate, progress milestones).
    \item Target encouragement at key journey points (1/2, 3/4).
  \end{itemize} \\

\hline

\multirow{4}{*}{\textbf{\parbox{2.2cm}{3. Frequency \& Control: Sparse, Valuable, and User-Directed}}} 
& \textbf{3.1 Sparse but High-Value Interactions} 
& \begin{itemize}[leftmargin=*,nosep,topsep=0pt,itemsep=0pt]
    \item ``The right frequency is a prompt every once in a while... too frequent is disruptive, and too infrequent weakens engagement.'' (P2)
    \item ``Prompts are better at key points like 1/2 or 3/4 of the way. They don't interfere with my walking.'' (P3, P6)
    \item ``The current frequency is a bit low for me; it could be higher.'' (P5)
  \end{itemize}
& \begin{itemize}[leftmargin=*,nosep,topsep=0pt,itemsep=0pt]
    \item \textbf{Adopt a 'less is more' philosophy for notifications.}
    \item Establish default frequency based on time or distance intervals.
    \item Focus on quality and relevance over quantity.
  \end{itemize} \\

\cline{2-4}
& \textbf{3.2 User Customization and Multi-modal Output} 
& \begin{itemize}[leftmargin=*,nosep,topsep=0pt,itemsep=0pt]
    \item ``If possible, I would like to customize the tone.'' (P1, P4)
    \item ``I hope it can have a voice version so I don't have to look down at my phone.'' (P3)
    \item ``There should be a 'snooze notifications' or 'mute this trip' button.'' (P7, P10, P11)
  \end{itemize}
& \begin{itemize}[leftmargin=*,nosep,topsep=0pt,itemsep=0pt]
    \item \textbf{Provide granular user control over agent behavior.}
    \item Implement basic controls (e.g., ``Pause reminders for 10 min'').
    \item \textbf{Develop voice-based output} to reduce physical interaction.
  \end{itemize} \\

\hline
\end{tabularx}
\end{table*}

\section{Discussion}
\label{sec:rq2}

Our crossover LMMs indicate that the Info-Motive condition produced reliably higher subjective outcomes than Information-only. For the primary outcome positive feelings, Info-Motive yielded a significant advantage (Table~\ref{tab:lmm_positive_feelings}: $\beta=-0.806$, $p<.001$), corresponding to a mean increase of $+1.03$ points on the composite and a large effect size ($d=1.85$) under conventional benchmarks~\cite{Cohen1988}. For the mechanistic outcome on user experience, Info-Motive likewise outperformed Info-Only (Table~\ref{tab:lmm_user_experience}: $\beta=-0.583$, $p=.007$, $d=1.46$). Together with the non-significant treatment$\times$sequence term in both models, these results support a robust within-subject advantage without detectable carryover contamination~\cite{Senn2002}.

These quantitative benefits are strongly supported by our thematic extraction of participant responses (Table~\ref{tab:thematic_analysis_results}), which uncovered predictable patterns consistent with the quantitative results. Participants greatly preferred companion-like supportiveness in the Info-Motive condition, terming it ``more like a real person'' and ``emotionally motivating'' (P2, P7). In stark contrast, they showed distaste for solely informative tones, labeling the Information-only condition ``cold and machine-like'' (P7) as well as ``without emotional engagement'' (P9). This relational break with the Information-only condition directly corresponds to the latter's worse performance on both positive feelings and user experience.

The thematic analysis also revealed the underlying mechanisms behind the effects through the unearthing of context-aware timing as an important success factor. The participants regularly reported that with timely motivational messages -- specifically during points of lethargy or important milestones of progress -- their experience was greatly improved and their motivation to persevere was enhanced. As one participant put it, ``When I was 3/4 through and was going to give up, its words of encouragement helped me persevere'' (P3). This result is direct confirmation of the JITAI principle behind our design for an intervention.

\rebadd{Compared to stationary chatbot usage, SmartWalkCoach introduces a dual-task interaction context where conversational support must compete with movement, navigation, and safety demands.
This explains why participants framed frequent or poorly timed prompts as burdensome--``too frequent is disruptive'' (P2).
Relatedly, interventions should be suppressed during high cognitive-load moments where participants' attention should be focused on their surroundings (e.g., street crossings). As one participant noted, ``In noisy environments or when crossing the street, I would ignore the prompts'' (P3, P7, P8). 
Therefore, an in-walk agent should be able to detect context and time its messages appropriately, in order to reduce both perceived burden and attentional disruption.}

The gains observed also match behavioral theory: adding briefer, context-specific encouragement addresses kindred needs suggested by Self-Determination Theory (competence, relatedness), thereby buoying intrinsic motivation and affect~\cite{DeciRyan2000}. From a UX point of view, affective appraisal is matched by perceptions of pragmatic/hedonic quality. As such, the improved user experience provides a viable means by which Info-Motive raises good feelings~\cite{Hassenzahl2020}. This also accords with the principles of JITAIs: providing support when apt occasions arise heightens acceptance and influence~\cite{Hardeman2019Systematic,Pielot2014Notifications,Mehrotra2016Receptivity}. \rebadd{Therefore, companionship agents should adaptively optimize interruption value density. In concrete terms, this could be achieved via implicit feedback loops: if prompts are repeatedly ignored, message frequency should be decreased; if users engage, it should be increased.}

Although our sample is small ($N{=}12$), the effect sizes are big and supplemented by quantitative controls (counterbalanced AB/BA experiment, carryover diagnostics) and qualitative coherence across participant reports, furnishing strong support for a positive answer to RQ2. Practically, integrated conclusions from these findings advocate for few-but-on-time motivational injections during the walk instead of continuous prompting. \rebdel{The thematic analysis particularly recommends the most effective delivery of interventions is during progress points (1/2, 3/4 points) and re-actively to behavioral signs of fatigue, but not during high-cognitive-load occurrences such as crossing streets.} Methodologically, future studies can formalize the assumed path by testing user experience as the mediator from condition to positive feelings (e.g., model-based causal mediation)~\cite{Imai2010Mediation}, and adding an offical satisfying item or UEQ-S subscale to better capture the ``satisfaction'' aspect under RQ2~\cite{Schrepp2017UEQS}. Nevertheless, our present evidence, which combines strong quantitative effects with deep qualitative observations, already significantly favors the conclusion that Info-Motive yields better motivation- and engagement-related outcomes compared to Info-Only in-the-wild.

\section{Limitations and Future Work}

Although SmartWalkCoach shows promising performance in delivering walking companioning with the power of AI, our study and system have some limitations that deserve discussion. The drawbacks also open up avenues for future development and study.

\subsection{Study Limitations}

The greatest weakness is our small diversity-poor sample size ($N=12$), where most participants were drawn from our University's student pool. Though adequate to uncover large effect sizes under our crossover design and in line with analogous HCI research~\cite{Sage2025Companionship}, our study would benefit from incorporating a larger and more varied participant pool to better generalize our results. Subsequent research must incorporate participants from mixed fitness levels, multiple age distributions, and mixed cultural backgrounds to test the cross-user segment robustness of our solution~\cite{Ng2022MobileHealthChildren}.

\rebadd{Moreover, while our study shows that our tool has a positive impact in motivating users to complete planned walks, it only captures short-term, session-level responses. To investigate the sustained impact of SmartWalkCoach on people’s lives (e.g., health benefits, exercise habits), a longer-term study is needed. Future work should therefore run a longitudinal deployment (e.g., months) to measure real usage trajectories (retention, habituation) and to examine whether a context-aware walking app with reflection and follow-up action planning meaningfully supports sustained walking habits over time.}

Another methodological weakness is our timely alerting mechanism. While our approach to virtual geofencing is promising, the existing segmenting strategy is mainly dependent on triggers based on distance. As identified under RQ3, even more refined timing models could utilize real-time physiological signals and contextual bandit models \cite{Lei2017ContextualBandits} to better recognize user states such as fatigue or stress. Existing work on active interactions for conversational assistants suggests that good timing is an optimal balance between user interruptibility and contextual feature relevance~\cite{TowardsProactiveInteractions2019}. Our existing system does not have such refined timing functionalities, thereby potentially overlooking points of optimal intervention.

\subsection{Technical Limitations}

The ability of our system's path adaptation remains limited. Although \emph{GeographyAgent} computes good initial paths, it is unable to re-plan paths on the fly during the walking phase based on the current user's state or environmental conditions. This is unlike better-starred navigation systems that are able to optimize paths constantly by utilizing live contextual information~\cite{Venkatraman2021RealTimeRouting}. Future deployments should include such dynamical replanning, achieving genuinely adaptive walking that is closer still to the paradigm of conversational navigation~\cite{FromLinksToDialogue2020}.

Another obvious weakness in our POI retrieval approach is shared by both \emph{GeographyAgent} and \emph{AccompanyAgent}. Both agents depend mostly on text-based reasoning by the LLM for producing and selecting POIs without any deep comprehension of user interests and inclinations. Although \emph{GeographyAgent} enjoys the advantage of direct conversational context for making selections, \emph{AccompanyAgent} is only able to work with very limited contextual information within walking sessions. This leads to POI suggestions that are logically suitable but may either not exactly comply with individual users' inclinations or fail to account for subtle situational considerations. Future work should be able to eliminate this by embedding explicit modeling of users' preferences, past interaction studies, and contextual awareness in the run-time to make the suggested POIs better personalized and situational.

Our approach to real-time fatigue detection presents a significant limitation during the in-walk stage. The triggering condition for the fatigue intervention is primarily based on a substantial drop in the user's walking speed. However, this heuristic is unreliable, as it cannot distinguish between true fatigue and other common urban walking scenarios, such as waiting for traffic lights, browsing shop windows, or simply pausing to enjoy the view. This lack of contextual discrimination leads to mistimed and disruptive interventions during our in-the-wild study. For instance, a significant proportion of participants reported that the encouragement was triggered while they were waiting at a crosswalk, which felt abrupt and ill-suited to the situation. Relying solely on speed decay without a richer understanding of the user's context and intent remains a key weakness of the current system. \rebadd{Additionally, the system does not currently send a message to motivate users to resume walking after a break.}

Furthermore, the interaction modality in our implementation combines visual and text-based elements~\cite{Li2025VisualCodeMooc}. This presents an accessibility issue as it demands visual attention that may sacrifice situational awareness when walking. Prior studies on the behavior of navigation systems suggest that visual interfaces have a negative influence on acquiring spatial knowledge and awareness of the environment~\cite{Brugger2019NavigationInfluence}. Future implementations of SmartWalkCoach should therefore consider incorporating voice-based conversational interactive navigation systems~\cite{Kaniwa2024ChitChatGuide}, providing users with the possibility of eyes-free and hands-free interaction.

Finally, our technology is limited by the depth of personalization. The system has built-in low-level user preferences but does not have mechanisms for long-term adaptation that develop over multiple uses. The inclusion of reinforcement learning methods analogous to those suggested for optimization of JITAIs~\cite{PersonalizedHeartSteps2019,Hochberg2016RLDiabetes} would allow the system to adapt its prompting strategy according to the response patterns observed throughout multiple walking sessions.

\subsection{Future Research Directions}

Based on these limitations, we outline the following potential future work avenues:
\begin{itemize}
    \item \textbf{Multi-modal activity monitoring for activity evaluation}: Prospective devices may tap the full potential of current smartphones and wearables to better estimate activity type, intensity, and form. Combining data from accelerometers, gyroscopes, and barometers might allow for walking patterns to be identified even in real time, so that more contextual feedback is possible~\cite{Bharti2018Human}.
    \item \textbf{Forward-looking modalities for external application}: Investigating voice-first modalities and audio-based acknowledgment is paramount for situational awareness during external activities. Research must consider evaluating hands-free alerts and full-duplex voice communication that reduces cognitive load while upholding user security under changing acoustics settings~\cite{Bokolo2025}.
    \item \textbf{Long-term adaptation for behavior retention}: Another key challenge in physical activity technologies is retaining activity over the long term. High-volume studies such as the NHS Active 10 application review illustrate the necessity of long-term compliance for substantial health benefits~\cite{Yerrakalva2025Active10}. Adapting reinforcement learning algorithms to tailor coaching approaches according to long-term user activity trends may fight attrition and reinforce habit formation~\cite{PersonalizedHeartSteps2019,Lei2017ContextualBandits,Hochberg2016RLDiabetes}. \rebadd{To achieve this, we will also need to train these on diverse user populations, including older adults, people with varying mobility or fitness levels, and diverse cultural contexts.}
    \item \textbf{Improved mechanisms to detect users' state}: Augmenting simple speed metrics with additional data streams could significantly improve the accuracy of state detection. For example, integrating data from the smartphone's inertial measurement unit could help classify user activity (e.g., walking, standing, running)~\cite{Bharti2018Human}. Furthermore, cross-referencing the user's live location with map API data could indicate if they are near a traffic intersection, a bus stop, or a POI, providing crucial contextual clues. By fusing movement data, geographical context, and potentially even user self-reports over time, a more robust model can be built to discern true fatigue from transient stops, ensuring that motivational support is delivered at genuinely appropriate moments.
    \item \rebadd{\textbf{Alternative segmentation mechanisms}: Future work could combine our conversational walking companion with other segmentation mechanisms that leverage meaningful spatial structures beyond distance thresholds. For example, geofences could be aligned with administrative or neighbourhood boundaries to reflect socio-spatial contexts that influence pedestrian behaviour~\cite{Gu2019Geofence}. Additionally, segmentation around key crossroads or intersection nodes can capture natural decision points in pedestrian networks where cognitive context shifts occur. Finally, defining geofences based on physical geographic features such as parks, waterfronts, or land-use transitions may better align with perceptual and activity boundaries in urban space~\cite{Shevchenko2024Geofencing}.}
    \item \textbf{Expansion into wider physical activity settings}: Although concentrating on walking, our architecture also seems naturally applicable to other similar settings such as walking interventions for weight loss~\cite{Walsh2016mhealth}, hiking trail recommendations, paced running workouts with pace mentoring, or active tourism where navigation is combined with cultural and historical data. Elements of gamification could also be added, built on evidence from mass-scale walking campaigns regarding what drives continued participation~\cite{Shameli2017GamificationWalking}.
\end{itemize}

\section{Conclusion}

SmartWalkCoach demonstrates that a relational, context-aware conversational companion can \rebrep{measurably enhance the lived experience of everyday walking when providing}{\ significantly increase participants’ positive feelings and enhance their walking experience through} integrated end-to-end support across pre-walk planning, in-walk accompaniment, and post-walk reflection. In a counterbalanced study with twelve participants, Information+Motivation produced large, statistically robust gains in participants' positive feelings and a composite user-experience index relative to an Information-only control, with no detectable carryover. Qualitative interviews explain these effects: people favored humanized and encouraging language, valued sparse but well-timed interventions at moments of fatigue or milestones, and rejected prompts that appeared during safety- or attention-critical contexts. Together, these findings answer our research questions by showing how narrowing agent roles and delegating geospatial computation reduces cognitive load in planning; that adding contextual motivation to otherwise equivalent guidance materially improves affect and experience; and that expression, timing, and frequency must be tuned to interruptibility and user control. 

We outline opportunities to address existing limitations via multi-sensor state estimation, contextual bandits or reinforcement learning for prompt policy optimization, dynamic replanning that accounts for terrain and amenities, and voice-first interaction to minimize visual demand. Beyond walking, the same agentic pattern may extend to running, active tourism, and outdoor well-being use cases where timely coaching must coexist with navigation. By coupling just-in-time motivational dialogue with strict attention to context and control, SmartWalkCoach points toward user interfaces that work with, not against, the rhythms of real-world movement.

\section*{GenAl Usage Disclosure}
The system described in this paper uses gpt-4o as all agents core LLM model within the app. GPT-5 assisted in handling non-novel boilerplate code in the data loader and drawing tools, as well as to assist in app development and write non-novel boilerplate code. All logic algorithms and function chains were built by the authors, and all AI-assisted code was verified by human reviewers. All Al outputs were reviewed, tested, and editedby the authors to ensure factual accuracy and reproducibility, following ACM's GenAl policy.

\bibliographystyle{ACM-Reference-Format}
\bibliography{bibliography}

\end{document}